\pdfoutput=1
\documentclass{PoS}

\newcommand{\ie}[0]   {\textit{i.e.}}
\newcommand{\eg}[0]   {\textit{e.g.}}

\newcommand{\mm}[1]   {\mathrm{#1}}

\newcommand{\DO}[0]   {D\O}
\newcommand\WINHAC[0] {\textsf{WINHAC}}
\newcommand\ZINHAC[0] {\textsf{ZINHAC}}

\newcommand{\Min}     {\mathrm{min}}
\newcommand{\Max}     {\mathrm{max}}

\newcommand{\flatDfDx}[2] {d\,{#1}/d\,{#2}}

\newcommand{\percent}{\,\%}

\newcommand{\MeV}    {\,\mathrm{MeV}}
\newcommand{\GeV}    {\,\mathrm{GeV}}
\newcommand{\TeV}    {\,\mathrm{TeV}}

\newcommand{\MW}     {{M_W}}

\newcommand{\Wp}     {{W^+}}
\newcommand{\BFWp}   {\mathbf{\Wp}}

\newcommand{\MWp}    {{M_\Wp}}

\newcommand{\Wm}     {{W^-}}   
\newcommand{\BFWm}   {\mathbf{\Wm}}

\newcommand{\MWm}    {{M_\Wm}}

\newcommand{\tolnu}  {\to l\,\nu_l}

\newcommand{\MZ}     {{M_Z}}

\newcommand{\lm}     {{l^-}}
\newcommand{\lp}     {{l^+}}

\newcommand{\Em}     {{e^-}} 
\newcommand{\ep}     {{e^+}}

\newcommand{\nue}    {{\nu_e}}

\newcommand{\mum}    {{\mu^-}}
\newcommand{\mup}    {{\mu^+}}

\newcommand{\numu}   {{\nu_\mu}}

\newcommand{\pTW}    {{p_{T,W}}}

\newcommand{\pTl}    {{p_{T,l}}}

\newcommand{\etal}   {{\eta_l}}

\newcommand{\val}     {{\mathrm{(v)}}}

\newcommand{\smartpap}{{p\hskip-7pt\hbox{$^{^{(\!-\!)}}$}}}
\newcommand{\ppbar}   {{p\,\pbar}}
\newcommand{\pbar}    {{\bar p}}
\newcommand{\pp}      {{p\,p}}

\newcommand{\kT}      {{k_T}}

\newcommand{\Asym}[1]   {\mathrm{Asym}^{(+,-)}\left(#1\right)}

\title{Prospect for precision measurements of
   $\mathbf{M_{W^+} - M_{W^-}}$  \& $\mathbf{M_W}$ at the LHC:
         Shortcuts revisited.
}

\ShortTitle{Measurement of the $W$ boson mass at the LHC: Shortcuts revisited.}

\author{\speaker{Florent Fayette}$^{,\;a}$, Mieczys{\l}aw Witold Krasny$^a$,
Wies{\l}aw P{\l}aczek$^b$, Andrzej Si\'odmok$^{b,a}$\\
\llap{$^a$}LPNHE, Pierre et Marie Curie Universit\'es Paris VI et Paris VII,\\
Tour 33, Rez-de-Chauss\'ee, 4 place Jussieu, 75005 -- Paris, France.\\
\llap{$^b$}Marian Smoluchowski Institute of Physics, Jagiellonian University,\\
Ulica Reymonta 4, 30-059 -- Cracow, Poland.\\
  E-mail: \email{fayette@lpnhep.in2p3.fr}, 
  \email{krasny@lpnhep.in2p3.fr},
  \email{wieslaw.placzek@uj.edu.pl},
  \email{andrzej.siodmok@cern.ch}}

\abstract{
The claim that the $W$ boson mass might be measured at the LHC with a precision of $\mathcal O(10\MeV)$ 
is critically reviewed. 
It is argued that such a precision cannot be achieved, unless a dedicated measurement program, specific 
to the LHC is pursued. We propose such a program. Its main target is to significantly improve 
the experimental control of the relative polarisation of the $W^+$, $W^-$ and $Z$ bosons. We propose to achieve this 
goal either by running dedicated isoscalar beams at the LHC or by running, in parallel to the 
standard $\pp$ collision program, a dedicated muon scattering ``LHC-support-experiment'' at the CERN SPS. 
One of these auxiliary measurements is necessary for the ``precision measurement program'' at the LHC, 
but not sufficient. It must be followed by dedicated measurement strategies which are robust with respect 
to both the systematic measurement uncertainties and to the perturbative and non-perturbative QCD effects. 
We propose such strategies and evaluate their precision. At the LHC, contrary to the Tevatron case, both 
the masses of the $W^+$ and of the $W^-$ bosons must be measured with high precision. 
In this context, we propose and evaluate LHC dedicated strategies to measure the difference of the masses of the 
$W^+$ and $W^-$ bosons and of the absolute mass of the $W$ boson assuming both masses are equal.
We show how one can overcome the obstacles in measuring the masses of $W^+$ and $W^-$ 
to a precision of $10\,\mathrm{MeV}$. 
We present a detailed evaluation of the precision of the proposed methods 
based on the studies of a large, $\mathcal O(10^{11})$, sample of simulated $W$ and $Z$ production events.}

\FullConference{The 2009 Europhysics Conference on High Energy Physics,\\
		 July 16--22 2009,\\
		 Krak\'ow, Poland.}

\begin{document}

\section{Introduction: the Tevatron results and the expectations for the LHC}
In hadronic collisions, the properties of the $W$ boson are extracted from its study in Drell--Yan like processes,
\ie{} $p\,\smartpap\to W+X\tolnu+X$ with $l\equiv\{e,\,\mu\}$.
The latest measurements of the mass of the $W$ boson from the CDF and \DO{} experiments~\cite{Aaltonen:2007ps}
are respectively $\MW=80.413\,\pm\,0.048\GeV$ and $\MW=80.401\,\pm\,0.043\GeV$.
These precisions measurements are facilitated in $\ppbar$ collisions because the dynamics producing 
the $\Wp$ and $\Wm$ bosons differ only by a $CP$ transformation. 
This leads eventually to a perfect charge symmetry in the transverse kinematics
of the leptons which are used in the extraction procedure of $\MW$.
A perfect illustration of the benefits gained over this symmetry is shown when both charged states are 
considered separately, \ie{} for the measurement of $\MWp-\MWm$.
In this context, the last CDF result is $\MWp-\MWm\approx 0.26\,\pm\,0.15\GeV$~\cite{Aaltonen:2007ps}.

Soon, the LHC should provide such important $W$ boson data that physicists will be left with the task to reduce 
the Systematic Errors (SE) coming from the uncertainties of both phenomenological and apparatus models.
Both ATLAS and CMS collaborations expect to control these two aspects well enough to measure $\MW$
with a precision of $\mathcal O(10\MeV)$~\cite{Besson:2008zs} for an integrated luminosity of $L=10\,\mm{fb}^{-1}$.
Nonetheless, the prospects that lead to such values did not explore the consequences of the LHC original context:
in $\pp$ collisions, there is no $CP$ symmetry between the dynamics of production of the $\Wp$ and $\Wm$ bosons.
Aware of this, we propose a program to measure the $W$ boson mass charge asymmetry 
$\MWp-\MWm$~\cite{Fayette:2008wt} and absolute mass $\MW$~\cite{Upcoming_MW} 
through the study of the transverse momenta of the charged leptons $\lp$ and $\lm$.

In the rest, we briefly justify why, in our view, both $\MWp-\MWm$ and $\MW$ measurements must be considered at the LHC
for a relevant precision measurement of $\MW$. After that, we present a re-evaluation of the size 
of the main SE along with possible solutions to decrease them using, each time,
dedicated systematic-robust strategies and observables.
Below, the charge asymmetry of an observable $x$ is defined as:
$\Asym{x}\equiv (\flatDfDx{\sigma}{x_\lp}-\flatDfDx{\sigma}{x_\lm})/(\flatDfDx{\sigma}{x_\lp}+\flatDfDx{\sigma}{x_\lm})$.
Details of our work can be found in our published papers and, more exhaustively, in the Ph.D. dissertations 
of Florent Fayette~\cite{Fayette:2008wt} and Andrzej Si\'odmok~\cite{Upcoming_MW}.

\section{Production of $\BFWp$ and $\BFWm$ bosons at the LHC in Drell--Yan like processes}
Here, like in our presentation~\cite{Fayette:EPS2009talk}, we review at the generator level and 
for both $\ppbar$ and $\pp$ collisions the behaviour of the most important kinematics entering 
in the extraction process of $\MW$.

At the level of the production, we focus on the transverse momenta distributions $\pTW$ of the $\Wp$ and $\Wm$ bosons.
For $\ppbar$ collisions they are exactly the same, while for $\pp$ collisions charge asymmetries are observed. 
Because the $W$ boson transverse motion has an important influence over the leptons transverse motion it appears 
vital for it to be well under control.
For the $W$ bosons leptonic decay, we study the charged leptons transverse momenta $\pTl$.
Again, for $\ppbar$ collisions there are no charge asymmetries at this theoretical level,
while for $\pp$ collisions the charge asymmetries are larger than the one seen in the production.
In a nutshell, these LHC-specific final state charge asymmetries result from the combination of the following effects:
(1) the excess, through the valence quarks, of matter over anti-matter, 
(2) the transverse motion of the colliding quarks (hence of the $W$) and eventually
(3) the $V-A$ coupling of fermions in electroweak interactions that magnifies the 
QCD initial state asymmetries from the items (1) and (2).

This loss of symmetry, with respect to the Tevatron, will open the door to large contributions to the error 
on $\MW$, in particular due to all the uncertainties tainting QCD.
Then, two solutions are possible: (1) $\Wp$ and $\Wm$ will have to be treated as distinct particles or 
(2), assuming $\MWp=\MWm$, the $W$ mass can still be extracted by considering both charged lepton data in the analysis, 
\ie{} $\MW=\frac{1}{2}(\MWp+\MWm)$, but, parallel to that, it becomes mandatory to consider a precision measurement 
of $\MWp-\MWm$ as well. This second solution is the one we choose.

\section{Expected systematic errors and strategies to reduce them}
As a consequence of the previous development, we addressed, on top of the usual sources of SE, 
new ones specific to the LHC original context.
For the apparatus we studied the impact from the miscalibration of the charged leptons absolute and relative
Energy Scale (ES) and track parameters resolution.
For the phenomenology, we considered the impact from the global uncertainties in the Parton Distributions Functions
(PDFs) and from the parametrisation of the intrinsic transverse momenta of the partons ${\kT}$.
In addition, we considered also new sources of SE coming from the uncertainties of specific sectors of the PDFs: 
(1) the $u^\val-d^\val$ valence (v) asymmetry,
(2) the $s-c$ asymmetry and
(3) the $b$ flavour (for $\MW$ only).
In the rest, like in our talk~\cite{Fayette:EPS2009talk}, we focus on the ES, $u^\val-d^\val$ and $s-c$ asymmetries
and choose only a few values to illustrate our discussion.

The study is realised by simulating with \WINHAC{} and \ZINHAC{} Monte Carlo~\cite{Placzek2003zg}
the expected statistic of $W\to e\,\nue,\;\mu\,\numu$ and $Z\to \ep\Em,\;\mup\mum$ events, 
for $L=10\,\mm{fb}^{-1}$, with $\sqrt s = 14\TeV$ and by smearing these Monte Carlo events with the ATLAS tracker resolution.
Cuts are performed upon the charged leptons transverse momenta and pseudo-rapidity distributions, respectively:
$\pTl>20\GeV$ and $|\etal|<2.5$. The $W$ boson properties, and the impact of the biases upon them, are extracted by performing
binned maximum likelihood fits to $\pTl$-based observables.

Our first estimations of the size of the SE are obtained using the $\Asym{\pTl}$ observable 
for the extraction of $\MWp-\MWm$ and the ratio of $W$ over $Z$ events for the one of $\MW$.
Considering errors of $0.5\percent$ for the ES, we end up with SE on $\MWp-\MWm$ of
$-550\MeV$ for charged-incoherent biases, and on $\MW$ of $200\MeV$ for charged-coherent biases.
For the $u^\val-d^\val$ and $s-c$ asymmetries we consider errors that would unlikely affect the $Z$ 
rapidity distribution.
Thus, in the valence sector, imposing $u^\val_\Max=1.05\,u^\val$ and $d^\val_\Min=d^\val-0.05\,u^\val$ 
we observe SE of $130\MeV$ ($80\MeV$) for $\MWp-\MWm$ ($\MW$).
For the $s-c$ sector, with $c_\Max=1.1\,c$ and $s_\Min=s-0.1\,c$ we observe SE of $-15\MeV$ ($-100\MeV$) 
for $\MWp-\MWm$ ($\MW$).

For the $\MWp-\MWm$ measurement, the errors from the absolute ES miscalibration could be reduced 
to $5\MeV$ by collecting half of the data with an inverted magnetic field in the tracker, 
so that both positive and negative tracks would be affected by the same biases.
For PDFs, the influence of the valence quarks can be reduced by restraining the analysed data to 
a narrow-central bin in the $W$ rapidity or, more realistic but less efficient, 
to the charged lepton pseudo-rapidity.
These prescriptions shows visible reduction on the size of the errors but, the $u^\val-d^\val$ 
uncertainty still contribute to the SE with $70\MeV$.
A perfect way to reduce greatly all those unknowns from the valence sector would be to collide isoscalar beams
which, in our study, for deuterium beams, decreased the error to $4\MeV$.
Still, some procedures invoked here are delicate and, if applicable, could be envisaged only in a more 
mature phase of the LHC program.
For that matter, alternative strategies with softer constraints on the LHC and its detectors will
be mentioned at the end.

For the $\MW$  measurement, the first improvement in the analysis is to enhance the QCD resemblance 
between the $W$ and the $Z$ data. 
In an earlier work~\cite{Upcoming_MW}, we proposed to trick the $W$ bosons to be produced by
partons with the same Bjorken $x$ as the one producing $Z$ bosons.
For that matter, a reduction of the nominal $\sqrt s$ by a factor $\frac{\MW}{\MZ}$ could be done 
when it would come to collect $W$ data.
In addition, a scaling of the magnetic field by a factor $\frac{\MW}{\MZ}$
would allow the charged leptons tracks from the $W$ to suffer from the same biases that 
affect the leptons tracks from the $Z$ bosons, and reduce potentially the impact of the relative ES miscalibration
to $5\MeV$.
In a next step, to account for the fact that $\MW \neq \MZ$, a QCD correction factor
was implemented and, although it lead to a reduction from the influence of $\kT$, it proved 
insufficient for the PDFs uncertainties.

To sum up, when the LHC data will be available we might face brick walls 
for a real enhancement of the precision on $\MW$ due to two main difficulties: 
(1) the calibration of the relative ES and 
(2) the uncertainties in specific sectors of the PDFs.
If we cannot free ourselves from these effects, 
we are left with the task to control them.
First, for the relative ES calibration, the $\Wp$ and $\Wm$ data,  being no longer suitable, 
could be substituted by the ``$Z^+$'' and ``$Z^-$'' data defined respectively as the 
positively and negatively charged leptons from the $Z$ boson decay.
Eventually a precision could be met to have impacts of $10\MeV$ for the errors on $\MWp-\MWm$
and $\MW$.
Second, for the PDFs, a dedicated muon deep inelastic scattering experiment
at the SPS collider~\cite{LOI:SPS_DIS} could be used to decrease the uncertainties from these
specific sectors that influence our measurements.
If the uncertainty on the $u^\val/d^\val$ ratio can reach $1\percent$, then the impact would be 
of $25\MeV$ ($12\MeV$) for $\MWp-\MWm$ ($\MW$).
Finally, if the $s-c$ asymmetry is controlled for $\MWp-\MWm$ ($\MW$) at $2\percent$ ($1\percent$),
the impact would be of $5\MeV$ ($10\MeV$).

\section{Conclusion}
Contrary to the earlier estimates, we conclude that one cannot reach a precision of $\mathcal O(10\MeV)$
for the measurement of the $W$ mass at the LHC unless a dedicated measurement program
\eg{} the one proposed in this contribution is pursued.

\end{document}